# АВТОМАТИЗИРОВАННОЕ ПОСТРОЕНИЕ СПИСКОВ СЕМАНТИЧЕСКИ БЛИЗКИХ СЛОВ НА ОСНОВЕ РЕЙТИНГА ТЕКСТОВ В КОРПУСЕ С ГИПЕРССЫЛКАМИ И КАТЕГОРИЯМИ

*А.А. Крижановский*
**Санкт-Петербургский институт информатики и автоматизации РАН**
*/ aka at iias dot spb dot su /*

В докладе представлены: алгоритм поиска синонимов (адаптированный HITS алгоритм), архитектура программы и оценка работы программы на тестовых примерах. Для тестирования алгоритма разработана программа Synarcher, выполняющая поиск синонимов (и близких по смыслу слов) в корпусе текстов специальной структуры (Википедиа). Результаты поиска представляются в виде графа с возможностью интерактивного поиска. Предложенное решение задачи поиска синонимов может использоваться при поиске информации (для расширения поисковых запросов), при составлении словарей синонимов.

## *Введение*

Увеличение числа и изменение качества электронных документов на локальных компьютерах и в сети Интернет позволяют адаптировать известные алгоритмы и предлагать новые для более точного поиска. Поиск похожих объектов (similarity search), кроме поиска похожих текстовых документов, включает задачу поиска семантически близких слов, задачу поиска похожих вершин графа. Для поиска синонимов и семантически близких слов применяют методы, учитывающие структуру гиперссылок, частоту словосочетаний и др.

В методах поиска, использующих структуру гиперссылок, учитываются весовые коэффициенты, назначенные каждому документу (в наборе документов с гиперссылками). Это позволяет вычислить относительную важность документа внутри данного набора (концепция авторитетных страниц). Алгоритмы HITS [Kleinberg, 1999] и PageRank (реализован в Google) предназначены для поиска интернет страниц, соответствующих запросу. Эти же алгоритмы позволяют искать похожие страницы (similar pages).

Метод извлечения контекстно связанных слов на основе частотности словосочетаний [Pantel, 2000] предлагается для поиска контекстно похожих слов (КПС) и для машинного перевода. Данными для поиска КПС служат (1) семантически близкие слова из тезауруса, (2) словосочетания из БД с указанием типа связи между словами. Для слова w формируется cohort w, т.е. группа слов, связанных одинаковыми отношениями со словом w, из базы словосочетаний. КПС слова w – это пересечение множества похожих слов (из тезауруса) с cohort w. Работа [Pantel, 2000] интересна формулами, предлагаемыми для вычисления сходства между группами слов.

В данной работе представлен адаптированный HITS алгоритм и его реализация в виде программной системы для поиска семантических синонимов в корпусе текстов с гиперссылками и категориями (Википедиа).

Трудности поиска синонимов определяются рядом причин. Во-первых, автору не известно общепринятой количественной меры для определения степени синонимичности значений слов. Можно утверждать, что одна пара слов более синонимична чем другая, но не ясен способ однозначно указать во сколько раз. Во-вторых, понятие синонимии определено не для слов, а для значений слов, т.е. синонимия неразрывно связана с контекстом. В-третьих, язык – это вечноизменяемая субстанция. Слова могут устаревать или получать новые значения. Особенно активное словообразование и присвоение новых значений словам наблюдается в науке, в её молодых, активно развивающихся направлениях.

Разработанные алгоритмы используют структуру ссылок в текстах, поэтому могут применяться к текстам на любом языке. Ссылки, связывающие страницы друг с другом, указываются экспертом. Предлагаемые алгоритмы применимы к корпусам текстов с гиперссылками и категориями. Эти тексты должны отвечать следующим условиям:
1) Каждому текстовому документу (статье) соответствует одно или несколько ключевых слов, отражающих содержание статьи. Например, в случае энциклопедии – энциклопедической статье соответствует одно слово – название статьи.
2) Статьи связаны ссылками. Для каждой статьи определены: набор исходящих ссылок (на статьи, которые упоминаются в данной статье) и входящих ссылок (на статьи, которые сами ссылаются на данную статью).
3) Каждая статья соотнесена одной или нескольким категориям (тематика статьи). Категории образуют дерево таким образом, что для каждой категории есть родитель-категория (кроме корня) и один или несколько детей-категорий (кроме листьев).

Данная структура не является абстрактным измышлением. Она имеет конкретное воплощение в структурах типа вики (wiki), получивших широкое распространение в последнее время в Интернете, например, в виде электронной онлайн энциклопедии

Википедиа (на русском, английском и других языках) (http://wikipedia.org). Анализ сетевого ресурса Википедиа представлен в [Holloway et al., 2005].

Алгоритмы, разработанные в системе, позволяют осуществлять поиск синонимов и близких по значению слов в английской и русской версии энциклопедии Википедиа. Причём нет принципиальных ограничений в применении программы к Википедиа на других языках, к вики ресурсам вообще и корпусам текстов, удовлетворяющих указанным выше требованиям.

### *Вики ресурсы*

Кратко осветим тему вики ресурсов, поскольку Википедиа является вики ресурсом. Перечислим значения слова. Вики (Wiki) – это тип сайтов, предоставляющих пользователям простой способ для добавления и изменения страниц сайта и, в особенности, предназначенных для совместной работы. Вики – это «простейшая онлайн база данных, которая, возможно, работает» [WhatIsWiki, 2005]. Также вики – это часть программного обеспечения (ПО) на стороне сервера, позволяющая пользователям создавать и редактировать содержание Интернет страниц с помощью любого Интернет броузера. Язык вики поддерживает гиперссылки (для создания ссылок между вики страницами) и является наглядным (по сравнению с HTML) и безопасным (нет JavaScript и Cascading Style Sheets) [Wiki, 2006].

Вики технологию изобрёл Ward Cunningham в 1995 г. Прелесть вики в том, что у каждой страницы есть ссылка "редактировать эту страницу". Пользователи могут редактировать страницы, читатели легко превращаются в писателей.

Чтобы вносить правки, пользователи регистрируются. У каждой страницы есть ссылка "страница истории". Ссылка ведёт на страницу, где указаны изменения, список авторов, упорядоченный по времени, и комментарии авторов к изменениям.

Концепция «свободного редактирования» имеет свои достоинства и недостатки. Открытость в редактировании текстов привлекает простых (технически не подкованных) пользователей, что позволяет развивать уже существующие вики ресурсы. К проблемам стоит отнести борьбу с вандализмом недружественных, а точнее, невежественных пользователей. Эта проблема решается благодаря возможности отката (в БД хранится история всех правок) и наличию истории страницы (указывается кто, что и когда правил). Решение об откате принимает администратор ресурса.

### *Тематическая связность авторитетных страниц*

Авторитетные страницы – это страницы, соответствующие запросу, имеющие больший удельный вес среди страниц данной тематики, т.е. большее число страниц ссылаются на данную страницу. Hub страница содержит много ссылок на авторитетные страницы.

Задача – извлечь тематически связанные авторитетные страницы из коллекции страниц. Простейший подход: упорядочить страницы по степени захода (число ссылок на страницу). Проблема такой простой схемы ранжирования в том, что могут быть найдены страницы с большой степенью захода, но тематически несвязанные.

Возможны следующие решения задачи:
1) *Решение HITS алгоритма*. Авторитетные страницы (по данной тематике) содержат не только большое число входных ссылок, но и пересекаются между собой. Это обеспечивается hub страницами, содержащими ссылки сразу на несколько авторитетных страниц (одной тематики). Т.о., научившись вычислять значения authority и hub страницы, получаем набор авторитетных страниц (тематически связанных), соответствующих запросу (здесь – слову, для которого ищем синонимы).
2) *Оригинальное решение* (на основе решения HITS алгоритма). В Википедии каждой странице эксперты приписывают несколько категорий. Категории образуют дерево, т.е. у каждой категории есть категория-родитель и несколько детей. Такая тематическая определённость страниц позволяет найденный список синонимов (с помощью адаптированного HITS алгоритма) разбить на кластеры, каждый из которых соответствует одному из значений исходного слова.

### *HITS Алгоритм*

В работе [Kleinberg, 1999] предлагается для поиска Интернет страниц (соответствующих запросу пользователя) использовать информацию, заложенную в гиперссылки. Демократическая природа Интернет позволяет использовать структуру ссылок как указатель значимости страниц (эта идея есть и в алгоритме PageRank). Страница p, ссылаясь на страницу q, полагает q авторитетной, стоящей ссылки. Для поиска существенно, что страница q соответствует тематике страницы p.

Каждой релевантной странице (найденной в алгоритме с помощью поискового сервера) сопоставляются веса *a* и *h*, которые показывают соответственно насколько страница является авторитетной и насколько она является хорошей hub страницей. В [Kleinberg, 1999] предлагаются следующие формулы для итеративного вычисления:

$$h_j = \sum_{i:(j,i)\in E} a_i \qquad (1)$$

$$a_j = \sum_{i:(i,j)\in E} h_i \qquad (2)$$

где $h_j$ и $a_j$ показывают, насколько страница *j* является хорошим указателем на релевантные страницы (т.е. hub) и насколько страница *j* является авторитетной страницей.

Таким образом, задачу поиска похожих Интернет страниц Kleinberg сводит к задаче поиска похожих вершин в графе на основе вычисления весов вершин.

### *Задача поиска похожих вершин в графе на основе весов hub и authority*

Дан направленный граф $G=(V, E)$, где $V$ – вершины (страницы), $E$ – дуги (ссылки), $v$ – вершина графа. Для каждой страницы $v$ известны два списка: $\Gamma^+(v)$ – это страницы, на которые ссылается данная статья, и $\Gamma^-(v)$ – это страницы, ссылающиеся на данную статью. Для каждой вершины определены значения двух весовых коэффициентов *authority* и *hub*: $\{v \in V: v_{authority} \in R, v_{hub} \in R\}$.

Необходимо найти набор вершин $A$, похожих на вершину $v$. Степень сходства (между $v$ и $A$) будет тем выше, чем больше будет hub-вершин $H$, указывающих и на $v$ и на вершины из $A$. При этом вершины $A$ являются авторитетными, т.е. на них указывают многие вершины той же тематической направленности, что и исходная вершина $v$.

Формализуем понятия: *похожие вершины* и *авторитетные вершины*, т.е. формализуем постановку задачи.

Необходимо найти множество вершин $A$, которые являются (i) *авторитетными* вершинами для исходной вершины $v$ (т.е. значение (3) велико относительно других подмножеств вершин той же мощности), (ii) *похожими* на исходную вершину $v$ в смысле наличия множества вершин $H$, ссылающихся одновременно и на исходную вершину $v$ и на вершины из $A$ (4). $H$ – это hub-вершины, т.е. значение (5) велико относительно других подмножеств вершин той же мощности. Задача – выбрать множество $A$ авторитетных вершин и $H$ hub вершин в значении (6), где $k$ – это один из параметров алгоритма.

$$A \subset V, |A|=N, \sum_{v \in A} v_{authority} \quad (3)$$

$$A \subset V, H \subset V, \forall a \in A \ \exists h \in H: \Gamma^+(h) \ni v, a \quad (4)$$

$$H \subset V, |H|=M, \sum_{v \in H} v_{hub} \quad (5)$$

$$A \subset V, H \subset V, k \in [0,1]:$$
$$k \cdot \sum_{v \in A} v_{authority} + (1-k) \cdot \sum_{v \in H} v_{hub} \to max \quad (6)$$

Назначение весов hub и authority вершинам-страницам может быть выполнено с помощью HITS алгоритма (формулы (1) и (2)).

### *Адаптированный HITS алгоритм с использованием методов кластеризации*

HITS алгоритм был адаптирован с учётом таких возможностей вики ресурсов, как: наличие двух списков страниц для каждой статьи (кто ссылается на данную статью $\Gamma^-(v)$, на кого ссылается данная статья $\Gamma^+(v)$), наличие у статей категорий, определяющих их тематическую принадлежность.

Адаптированный алгоритм поиска включает шаги:
1) В корневой набор включаются те страницы, на которые ссылается исходная страница (вместо поиска с помощью поискового сервера в оригинальном HITS алгоритме).
2) В базовый набор включаются страницы, которые связаны ссылками с корневым набором.
3) Итеративно вычисляются веса с помощью формул (1) и (2) для поиска авторитетных и hub страниц среди страниц базового набора.
4) Алгоритм кластеризации на основе категорий статей (шаг отсутствует в оригинальном алгоритме).

Предлагается следующий алгоритм кластеризации статей (шаг 4) на основе категорий статей. Определим структуры данных, используемые в алгоритме и представим псевдокод алгоритма.

*Переменные и структуры данных алгоритма кластеризации.*

- $G=(V, E)$ – направленный граф, где $V$ – вершины (статьи и категории, т.е. два типа вершин), $E$ – дуги (три типа дуг: между статьями, между категориями, между статьями и категориями).
- $E_{sorted}$ – массив рёбер, упорядоченный по весу ($E_{sorted}[0]$ – ребро с минимальным весом).

Clusters – список кластеров, которые нужно построить.

Для каждого ребра $e$ определены следующие поля:
- $e_{c1}$ и $e_{c2}$ – это указатели на два соединяемых кластера,
- $e_{weight}$ – вес ребра, равный суммарному весу объединяемых кластеров $c_1$ и $c_2$.

Для кластера-вершины $c$ определены такие поля:
- $|c_{edges}|$ – число объединённых рёбер кластера (рёбер между вершинами-категориями кластера),
- $|c_{articles}|$ – число статей, которые ссылаются на категории в кластере (знак мощности множества $|.|$ используется в силу его наглядности, однако, поскольку $|c_{articles}|$ – это переменная, то ей можно присваивать значение (см. напр. строку 5b в алгоритме),
- $c_{weight}$ – вес кластера.

*Входные параметры.*

*MaxClusterWeight* – максимально разрешённый вес кластера. При вычислении веса кластера учитываются: число статей в кластере, веса объединяемых кластеров.

Процесс кластеризации состоит из двух шагов: предобработка (инициализация массива кластеров, присвоение начальных значений полям вершин и рёбер) и сам алгоритм кластеризации.

*Предобработка*
(две косые черты '//' отделяют комментарий от псевдокода)

1. Построить кластеры (массив Clusters) по категориям: изначально каждый кластер соответствует отдельной вершине (категории). Приписать каждому кластеру (за счёт содержащихся в кластере категорий):
   a) $|c_{articles}|$ = число статей, которые ссылаются на категории в кластере,
   b) $c_{weight} = 1 + |c_{articles}|$    // изначально вес кластера – это число категорий в кластере (изначально одна категория) и число статей, которые ссылаются на эту одну категорию,
   c) $c_{category\_id}[0]$ = category$_{id}$    // присваиваем кластеру уникальный идентификатор id первой (и единственной пока) категории, добавленной в кластер (у каждой категории и статьи Википедиа есть уникальный идентификатор).
2. Для каждого ребра между категориями создать ребро между кластерами. Каждому ребру $e$, соединяющему два кластера c1 и c2 определить вес так:
   $e_{weight} = c1_{weight} + c2_{weight}$

*Алгоритм*
1. $E_{sorted} = sort(e_{weight})$;    // сортировка рёбер по весу
2. while($|E_{sorted}| > 0$ && ($E_{sorted}[0]$ < MaxClusterWeight)) BEGIN
3.    $e = E_{sorted}[0]$;    // $v_1$, $v_2$ – вершины смежные ребру e
4.    $E_{sorted} = E_{sorted} \setminus \Gamma(v_2)$;    // удалить из упорядоченного массива рёбер рёбра смежные $v_2$
5.    merge(e);    // объединить вершины-кластеры $v_1$ и $v_2$ в кластер $v_1$, т.е. добавить вершину $v_2$ в кластер $v_1$, изменив свойства $v_1$ так:
   a) $v_{1\ weight}\mathrel{+}= v_{2\ weight}$;    // увеличить размер кластера (число категорий и статей)
   b) $|v_{1\ articles}|\mathrel{+}= |v_{2\ articles}|$;    // увеличить число статей
   c) $|v_{1\ edges}|\mathrel{+}= |v_{2\ edges}|$;    // увеличить число рёбер
   d) $v_{1\ category\_id}[]\mathrel{+}= addUnique(v_{2\ category\_id}[])$; // добавили категории без повторов
6.    passEdges();    // все рёбра смежные вершине $v_2$ передать вершине $v_1$ (рёбра без повторений, это не мультиграф).
7.    Esorted = Esorted \ edge ($v_1$, $v_2$);// удалить ребро ($v_1$, $v_2$);
8.    updateEdgesOfMergedCluster;    // обновить указатели на вершины, удалить рёбра (вершины и рёбра, смежные удаляемой вершине)
9.    updateEdgeWeight($v_1$);    // пересчитать значения весов для всех рёбер смежных $v_1$
10.   remove(Clusters, $v_2$);    // удалить кластер $v_2$ из массива кластеров
11.   $E_{sorted} = sort(e_{weight})$    // пересортировка рёбер, сложность O(N), т.к. нужно обновить порядок только тех рёбер, которые смежны вершине $v_1$
12. END
13. Return Clusters

Результат работы алгоритма – это кластеры категорий (Clusters). По кластеру категорий получаем кластер статей.

Заметим, что одна статья может ссылаться на несколько категорий. Поэтому одна статья может принадлежать нескольким кластерам. Но категория принадлежит ровно одному кластеру.

*Реализация*

*Данные*. Система Synarcher работает с данными Википедиа в MySQL формате (структура таблиц соответствует требованиям системы MediaWiki - основа для работы Википедиа). Поиск синонимов проводился на основе данных английской и русской версии энциклопедии Википедиа. Вероятно, система Synarcher сможет искать синонимы и на других вики-ресурсах, основанных на MediaWiki.

*Требования к ПО*. Для запуска системы Synarcher на стороне клиента требуется Jave Runtime Environment (JRE) версии не ниже 1.3.0. Система тестировалась в операционных системах Windows XP и Mandrake Linux.

*Клиент-серверная архитектура*. Ресурсы Википедиа доступны благодаря слаженной работе на стороне сервера таких программ и сервисов, как: MySQL, Apache, php, MediaWiki. Общедоступный сервер Википедиа (http://en.wikipedia.org) не использовался, поскольку данная реализация поиска синонимов требует значительной вычислительной нагрузки БД. Поэтому обрабатывалась локально установленная база данных MySQL Википедиа.

*Интерактивный режим работы с графом*. Результаты поиска представлены в виде графа. Вершины соответствуют названиям статей энциклопедии (статья – это гипертекстовая страница), дуги указывают наличие гиперссылок между статьями. Пользователь может раскрыть вершину (отобразить список соседей), спрятать соседей, пометить вершину как синоним. Изначально пользователь вводит слово, и система выполняет автоматический поиск синонимов. Затем пользователь помечает (опция rate/unrate) те слова, которые действительно являются синонимами (работа в интерактивном режиме с таблицей или визуальным представлением графа). Данная клиент-серверная система настроена так, что параметры поиска для каждого искомого слова и список отобранных пользователем синонимов хранятся на компьютере пользователя.

Представление в виде графа результатов поиска синонимов основывается на программе TouchGraph WikiBrowser V1.02 (http://www.touchgraph.com). Экран делится на две части вертикальной полосой, с левой стороны можно задать параметры поиска, посмотреть энциклопедическую статью, соответствующую выбранному слову, посмотреть результаты в табличном и текстовом виде, с правой стороны представлена часть вики графа (Рис. 1).

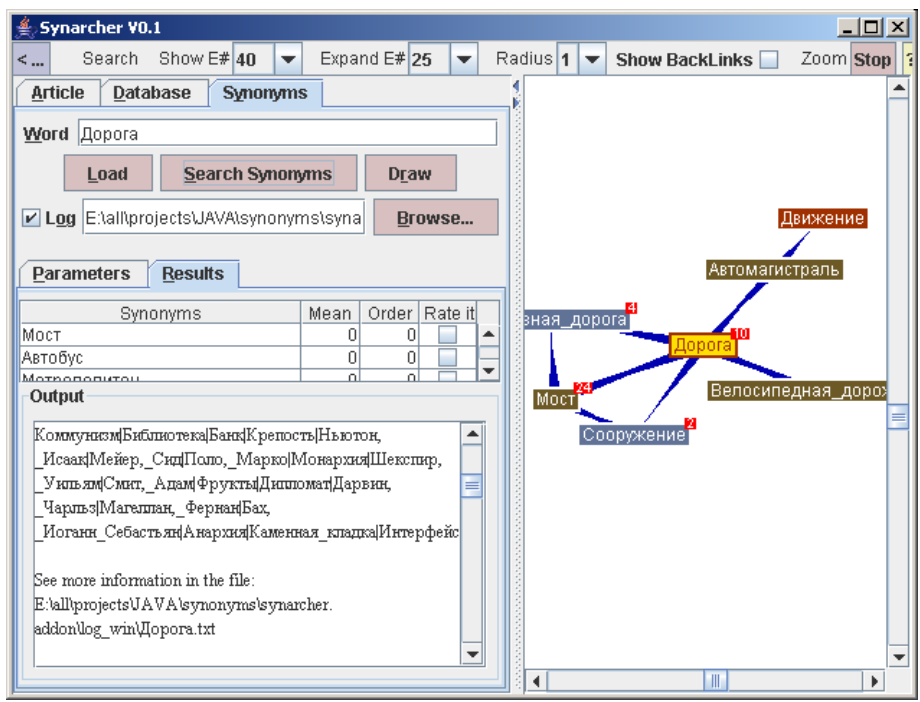

Рис. 1 Результаты поиска синонимов к слову "дорога" в виде таблицы, текста (поле Output) и графа

### *Эксперименты*

Эксперименты проводились с локальной версией английской энциклопедии Википедиа, соответствующей онлайн версии от 8 марта 2005. Английская Википедиа включала 901,861 страниц, 18,380,035 гиперссылок, русская – 30,161 страниц, 468,771 гиперссылок.

Нужно отметить, что полностью автоматизировать поиск синонимов не удалось, т.е. программа строит список слов, многие из которых не являются синонимами исходному. Необходим дополнительный интерактивный поиск (на графе и в таблице) с помощью эксперта. Таким образом, задача программы – это отбор, фильтрация потенциальных синонимов, которые служат сырьём для выбора эксперта.

В ходе интерактивного поиска с помощью программы Synarcher для слова *robot* экспертом были найдены следующие семь синонимов: *android, golem, homunculus, domotics, replicant, sentience, parahumans* (подчёркнуты синонимы, присутствующие в тезаурусах WordNet и Moby; значения слов см. http://en.wikipedia.org). WordNet 2.0 содержит только два синонима к слову *robot*. Это *automaton* и *golem*.

Для слова astronaut с помощью программы были выбраны четыре синонима: *cosmonaut, taikonaut, spationaut, space tourist*. WordNet предлагает два синонима: *spaceman, cosmonaut*. Moby Thesaurus 1.0 не содержит слова *robot*, а слову *astronaut* в нём соответствуют 79 слов, из которых шесть слов можно отнести к синонимам или словам близким по значению: *aeronaut, cosmonaut, pilot, rocket man, rocketeer, spaceman*.

Итак, для слов *astronaut* и *robot* было найдено два и шесть синонимов, отсутствующих в тезаурусах WordNet и Moby.

В виду небольшого размера энциклопедии Википедиа на русском языке, результаты поиска оставляют желать лучшего. В то же время стремительный рост русской Википедиа позволяет надеяться на интересные эксперименты в ближайшем будущем.

### *Заключение*

В работе представлен адаптированный HITS алгоритм для поиска синонимов и близких по смыслу слов в корпусах текстов с гиперссылками и категориями. Алгоритм реализован в программе Synarcher, осуществляющей поиск в английской и русской версиях энциклопедии Википедиа. Программа будет доступна по адресу

http://sourceforge.net/projects/synarcher.

Проведён ряд экспериментов, показывающих возможность успешного поиска синонимов с помощью данной программы. Для слов *robot* и *astronaut* были найдены синонимы, отсутствующие в тезаурусах WordNet и Moby. Это можно объяснить свойствами источника данных (энциклопедии Википедиа): наличие статей и на классическую для энциклопедии тематику (наука, искусство, политика и др.) и на самую современную тематику (база обновляется, буквально, каждый день), большое количество статей (в английской версии их число превысило размер энциклопедии Британника).

Предложенное решение задачи поиска синонимов может использоваться в поисковых системах (расширение запросов с помощью тезаурусов [Браславский, 2004]), в системах машинного перевода, при составлении словарей синонимов.

*Благодарности*